\documentstyle[11pt,newpasp,twoside, epsf]{article}
\def\edcomment#1{\iffalse\marginpar{\raggedright\sl#1\/}\else\relax\fi}
\reversemarginpar

\begin{document}
\title{Nonradial Pulsations in Classical Cepheids of the Magellanic Clouds}
\author{Pawe{\l} Moskalik}
\affil{Copernicus Astronomical Center, ul.\thinspace Bartycka 18, 00-716 Warsaw, Poland}
\author{Zbigniew Ko{\l}aczkowski}
\affil{Wroc{\l}aw University Observatory, ul.\thinspace Kopernika 11, 51-622 Wroc{\l}aw, Poland}
\author{Tomasz Mizerski}
\affil{Warsaw University Observatory, Al.\thinspace Ujazdowskie 4, 00-478 Warsaw, Poland}

\begin{abstract}
We have performed systematic frequency analysis of the LMC Cepheids observed by OGLE 
project. Several new types of pulsation behaviour are identified, including 
triple-mode and amplitude-modulated double-mode pulsations. In $\sim\! 10$\% of the 
first overtone Cepheids we find low amplitude secondary periodicities corresponding to 
nonradial modes. This is the first evidence for excitation of nonradial oscillations 
in Classical Cepheid variables. 
\end{abstract}

\section{Introduction}

For decades Cepheids and RR~Lyrae stars have been considered primary examples of 
purely radial pulsators. This simple picture has changed in recent years, when 
nonradial modes have been detected in substantial fraction of RR~Lyrae variables in 
various stellar systems (Kov\'acs 2002; Moskalik \& Poretti 2003; Alcock et al. 2003). 
Motivated by these findings, we have undertaken a systematic search for nonradial 
Fpulsators among Pop.~I Cepheids of the Magellanic Clouds. In this paper we present our 
results for the LMC, with particular emphasis on the new types of pulsation behaviour 
detected during the survey. The analysis of the SMC Cepheids is currently underway and 
will be presented at a later date.

\section{Search for multiperiodicity}

The primary source of data for our analysis is the OGLE-II photometry obtained with 
the Difference Image Analysis method (\.Zebru\'n et al. 2001). This photometry covers 
the timebase of $1000-1200$\thinspace days, with $250-500$ I-band flux measurements 
per star. Such a long and uniform dataset is particularly well suited for our task. 
The search for multiperiodicity is conducted for all LMC Cepheids identified by the 
OGLE team (Udalski et al. 1999; Soszy\'nski et al. 2000), except objects marked in 
their cataloge as FA, which are mostly Pop.~II variables. The search is performed with 
a standard consecutive prewhitening technique. First, the data is least square fitted 
with the single frequency Fourier sum of the form 

$$I(t) =  \langle I\rangle + \sum_{k} {\rm A}_k \cos (2\pi k{\rm f}_0 t + \phi_k)\eqno(1)$$

\noindent with pulsation frequency ${\rm f}_0$ being also optimised. The Fourier 
transform of residuals of the fit is then calculated over the range of $0\! -\! 
5$\thinspace c/d, in order to reveal any secondary periodicities. In the next step, a 
new Fourier fit with all frequencies identified so far and {\it their linear 
combinations} is performed and the fit residuals are searched for additional 
periodicities again. The process is repeated, until no new frequencies appear. The 
most interesting results of our survey are summarised below.

\section{FO/SO/TO triple-mode Cepheids}

In two of the previously known FO/SO double-mode Cepheids a third strong periodicity 
is detected. The stars are listed in Table 1. The period ratio of ${\rm P}_{\!  
3}/{\rm P}_{\! 2}\simeq 0.84$ identifies the new mode as a third radial overtone. With 
three radial modes observed, these stars will strongly constrain the Cepheids 
evolutionary tracks. Preliminary calculations show already that both objects are on 
the first crossing of the instability strip (Dziembowski, priv.~comm.). 

\begin{table}
\caption{Triple-mode Cepheids}
\tabcolsep=9pt
\begin{center}
\begin{tabular}{ccccccc}
\hline\\[-7pt] 
Star & & ${\rm P}_{\! 1}$\,[day] & ${\rm P}_{\! 2}$/${\rm P}_{\! 1}$ & ${\rm P}_{\! 3}$/${\rm P}_{\! 2}$ & ${\rm A}_{\! 2}$/${\rm A}_{\! 1}$ & ${\rm A}_{\! 3}$/${\rm A}_{\! 1}$ \\
[3pt] 
\hline\\
[-7pt]
LMC SC3--360128 & & 0.54128 & 0.8056 & 0.8400 & 0.167 & 0.311 \\ 
[1pt]
LMC SC5--338399 & & 0.57951 & 0.8052 & 0.8403 & 0.111 & 0.178 \\
[3pt]
\hline\\
[-20pt]
\end{tabular}
\end{center}
\end{table}

\section{FO/SO double-mode Cepheids with amplitude modulation}

In 18 out of 55 FO/SO double-mode Cepheids we detect residual signal in the vicinity 
of primary pulsation frequencies. With the OGLE-II data, these se\-condary peaks are 
always unresolved from the primary ones. Therefore, we repeat the analysis with the 
MACHO data (Allsman \& Axelrod 2001), which is of lower quality, but provides a much 
longer timebase ($\sim 2700$\thinspace day). In 8 stars we are now able to resolve the 
secondary power into two peaks located on opposite sides of the primary (radial) peak. 
Together with the primary frequency they form a close {\it equidistant triplet}. 
Typical example od such a pattern is displayed in Fig.\thinspace 1. The triplet 
structure can appear either around one (3 stars) or around both radial modes (5 
stars). In one object we detect a triplet and an {\it equally spaced quintuplet}. Weak 
evidence of a quintuplet structure is also noticable in two other stars. In all stars 
amplitudes of the sidepeaks are below $0\fm 03$. The separation of components is 
always very small and corresponds to the beat period of more than $750$\thinspace day. 
When present around both radial modes, the two triplets/quintuplets have {\it 
identical frequency spacings}. 

One more star can be resolved with MACHO data, and instead of triplets we find only 
doublets. In this particular object the secondary peaks are detected at very low 
signal-to-noice ratio. Therefore, it is likely that the detected doublets are in fact 
part of triplets, with the third component being burried in the noise. 

The triplet/quintuplet frequency structures discovered in the FO/SO double-mode 
Cepheids can result from excitation of nonradial modes (Nowakowski \& Dziembowski 
2001), but they can also be produced by periodic amplitude and phase modulation of the 
radial modes. The latter hypothesis seems to be more likely, as it can naturally 
explain why the frequency splittings of both radial modes are always the same. 

\begin{figure}
\plotfiddle{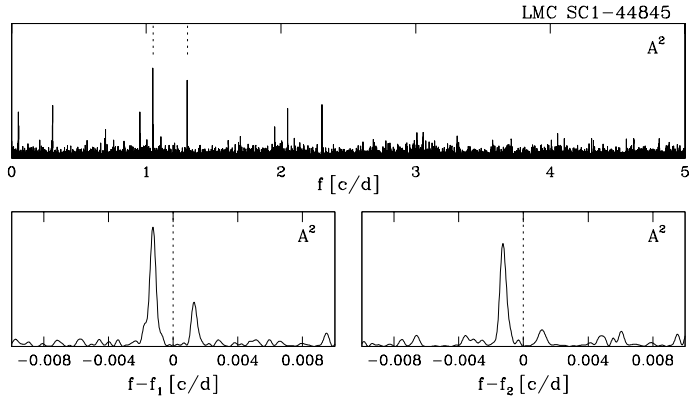}{46mm}{0.0}{58}{58}{-188}{-219} 
\caption{Power spectrum of FO/SO double-mode Cepheid LMC\thinspace SC1-44845 after 
prewhitening with two radial frequencies and their linear combinations. Lower panels 
display the fine structure around the radial modes. The frequencies of the (removed) 
radial modes are indicated by the vertical dashed lines.} 
\end{figure}

\section{Nonradial pulsators}

In 46 first overtone Cepheids we detect well resolved secondary peaks close to the 
dominant radial mode. Two examples of such a behaviour are shown in Fig.\thinspace 2. 
In most stars only one peak is present, but in several cases two or even three peaks 
are found, usually on the same side of the primary frequency. The observed patterns 
cannot be explained by amplitude and/or phase modulation. The period ratios are {\it 
incompatible} with those of radial modes, implying that the newly detected 
periodicities must correspond to {\it nonradial modes of oscillation}. 

The secondary peaks are always very small. Their amplitudes are $10-30$ times lower 
than the amplitude of the radial mode. In most cases nonradial modes are found on the 
low frequency side of the dominant peak. The frequency separation $\Delta{\rm f} = 
{\rm f}-{\rm f}_0$ is typically between $-0.15$\thinspace c/d and $+0.05$\thinspace 
c/d. This is very similar to $\Delta{\rm f}$ range found for the nonradially pulsating 
RRc stars (Alcock et al. 2000). 

46 stars in which nonradial modes are detected constitute 10\% of the LMC first 
overtone Cepheid sample. In sharp contrast, {\it nonradial modes are not found} in any 
of the 720 fundamental mode Cepheids, with a possible exception of LMC\thinspace 
SC17-157908, where marginal detection can be claimed. Clearly, the incidence rate on 
nonradial pulsations is dramatically lower in the fundamental mode variables.

Nonradial modes are also found close to the first overtone in two of the FU/FO 
double-mode Cepheids. In both stars, however, secondary peak is only weakly 
significant and requires confirmation with better data. 

\begin{figure}
\plotfiddle{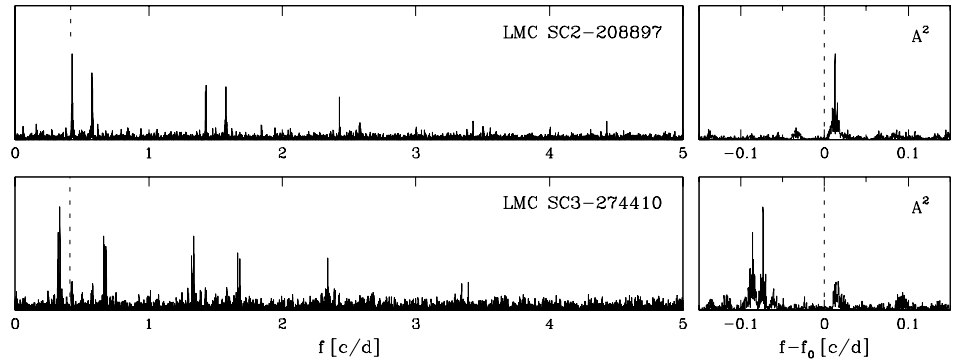}{43.5mm}{0.0}{115}{115}{-248}{-306} 
\caption{Power spectrum of two first overtone Cepheids after prewhitening with the 
primary frequency and its harmonics. The right column of the plot displays the fine 
structure in the vicinity of the radial mode. The frequency of the (removed) radial 
mode is indicated by the vertical dashed line.} 
\end{figure}

\vfill

\bigskip

\noindent{\bf Acknowledgements.} This work has been supported by Polish 
KBN grants 5~P03D~012~20 and 5~P03D~030~20.

\end{document}